%% file: main.tex
\documentclass[10pt,twocolumn,letterpaper]{article}

\usepackage[pagenumbers]{wacv} 

\input{preamble}

\definecolor{wacvblue}{rgb}{0.21,0.49,0.74}
\usepackage[pagebackref,breaklinks,colorlinks,allcolors=wacvblue]{hyperref}
\usepackage[accsupp]{axessibility}  
\newcommand{\Inputs}{\item[\textbf{Inputs:}]}
\newcommand{\Outputs}{\item[\textbf{Outputs:}]}

\def\wacvPaperID{714} 
\def\confName{WACV}
\def\confYear{2026}

\title{General and Domain-Specific Zero-shot Detection of Generated Images via Conditional Likelihood}

\author{
Roy Betser$^{1,2}$ \quad
Omer Hofman$^{2}$ \quad
Roman Vainshtein$^{2}$ \quad
Guy Gilboa$^{1}$\\
$^{1}$Technion - Israel Institute of Technology \quad
$^{2}$Fujitsu Research of Europe\\
{\tt\small roy.betser@fujitsu.com}
}

\begin{document}
\maketitle

\begin{abstract}
The rapid advancement of generative models, particularly diffusion-based methods, has significantly improved the realism of synthetic images. As new generative models continuously emerge, detecting generated images remains a critical challenge. While fully supervised, and few-shot methods have been proposed, maintaining an updated dataset is time-consuming and challenging. Consequently, zero-shot methods have gained increasing attention in recent years. We find that existing zero-shot methods often struggle to adapt to specific image domains, such as artistic images, limiting their real-world applicability. In this work, we introduce CLIDE, a novel zero-shot detection method based on conditional likelihood approximation. Our approach computes likelihoods conditioned on real images, enabling adaptation across diverse image domains. We extensively evaluate CLIDE, demonstrating SOTA performance on a large-scale general dataset and significantly outperform existing methods in domain-specific cases. These results demonstrate the robustness of our method and underscore the need of broad, domain-aware generalization for the AI-generated image detection task. Code is available at \url{https://tinyurl.com/clide-detector}.

\end{abstract}

\section{Introduction}


The rapid progression of generative methods, particularly diffusion models \citep{dhariwal2021diffusion, podell2023sdxl}, has made synthetic imagery remarkably realistic. These advancements boost creative and practical applications. However, they also pose significant risks in social media, journalism, and security, where manipulated imagery can facilitate misinformation, impersonation, and other forms of digital deception \cite{verge2023pope, Scott2023AIphoto}. Hence, the development of reliable detection methods is crucial to preserving public trust and ensuring digital integrity.

\begin{figure}[ht]
    \centering\includegraphics[width=0.45\textwidth]{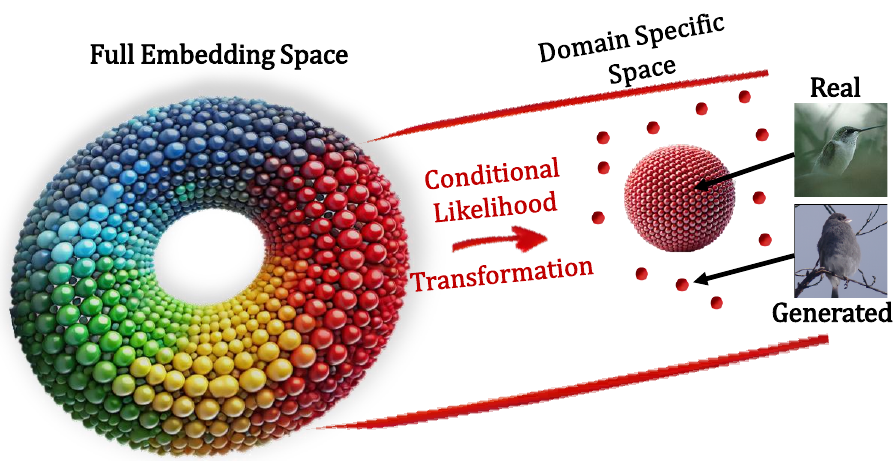}
    \caption{\textbf{Conditional likelihood illustration.}  Our adaptive image detection method.
    When conditioned by real images from a specific domain (red dots), generated images become unlikely.} 
    \label{fig: teaser}
\end{figure}

Prior research on generated image detection includes classification-based and semi-supervised approaches \citep{baraheem2023ai, bird2024cifake, cioni2024clip, qiao2024unsupervised}. The performance of these methods heavily depends on the quality and diversity of the training data, which includes both synthetic and real images \cite{epstein2023online}. However, as novel generative models continue to emerge at a rapid pace, maintaining an updated synthetic training set becomes not only impractical but sometimes impossible. This is especially true in misinformation campaigns orchestrated by actors who may develop or access entirely new generation techniques. This highlights the need for detection frameworks that can adapt to unseen generative techniques without relying on new data from every new model.


\begin{figure*}[ht]
    \centering\includegraphics[width=0.85\textwidth]{general_images_example.pdf}
    \caption{\textbf{General real and generated images.} Real and generated images of general real-world imagery. A blurred real image is misclassified as generated. A realistic generated image of an elephant is misclassified as real. All other examples are correctly classified.}
    \label{fig: cond example}
\end{figure*}

Zero-shot learning is well-suited for generated image detection as it eliminates the need for large training datasets, prior knowledge of generative models, or continual monitoring of new techniques. In this context, as established in \citep{ricker2024aeroblade, he2024rigid, cozzolino2024zero, brokman2025manifold}, zero-shot means distinguishing between real and generated images without any exposure to generated content or task-specific training. While this line of work has enabled promising progress, existing zero-shot methods still exhibit limited robustness and sharp performance swings across generators \cite{brokman2025manifold}.

To meet real-world needs, we formalize \emph{domain-specific zero-shot detection}: operate within a narrow target domain using only a set of \emph{real} domain image, with no access to generated samples and no task training (see Supp. Sec.~I) for task definition and discussion). This setting, analogous to cross-task robustness in LLM-generated text detection \cite{wu2025survey}, reveals severe failures of existing image detectors on narrow domains. We introduce two practical benchmarks (damaged cars, invoices) and show that CLIDE
sustains high accuracy where prior methods degrade.

Our fundamental premise is that foundation models, trained on huge datasets of real images, capture well the subtle attributes differentiating real images from generated ones. 
However, accessing likelihood in such black-box models is a very complex task, and more broadly, computing likelihoods for images is a challenging task in computer vision \citep{gao2023implicit, kawar2023imagic}. Current computer vision models primarily rely on likelihood gradients or score functions, which constrain direct likelihood computation \citep{ho2020denoising}.
A recent study \cite{betser2025whitened} proposed a simple method to estimate image likelihood based on Contrastive Language-Image Pre-training (CLIP) \cite{radford2021learning}. CLIP is a foundation model that 
is widely used across various computer vision tasks \cite{mokady2021clipcap, cioni2024clip}. We generalize this approach by introducing a conditional likelihood framework that enables domain adaptation and strengthens robustness for zero-shot generated image detection. 

\noindent \textbf{Our main contributions are:}
\begin{enumerate}
    \item We introduce a conditional likelihood approximation model for images based on CLIP embeddings, and show how the model can be validated through statistical tests. 
    \item We propose CLIDE, a zero-shot generated image detection method applicable to both general images (\cref{fig: cond example}) and domain-specific images (\cref{fig: teaser}). CLIDE significantly outperforms four state-of-the-art zero-shot detection baselines on a large-scale general image dataset.
    
    \item We demonstrate the domain adaptation capabilities of CLIDE on three specific domains: artistic images, damaged car images and invoice document images. 
    Our method consistently maintains high performance, while other methods experience a decrease in effectiveness in domain-specific content. 
    \item We present two new synthetic datasets of damaged cars (6K images) and invoice images (3K).
\end{enumerate}

\section{Background and Related Work}

\textbf{Generated image detection.}
Early work on detecting AI-generated images predominantly employed supervised methods, often training standard CNNs to distinguish real and generated images \citep{wang2020cnn,baraheem2023ai,epstein2023online,bird2024cifake}. 
Subsequent studies refined these methods by focusing on key phenomenological cues to enhance separability \citep{bammey2023synthbuster,martin2023detection,zhong2023rich,wang2023dire,tan2024data,chen2024single}. However, they largely relied on comprehensive datasets from known generative models, limiting their ability to manage newly emerging generative techniques \citep{ojha2023towards}.

To address the need for broader generalization, unsupervised and semi-supervised techniques were explored \citep{zhang2022exposing,zhang2022improving,qiao2024unsupervised,cioni2024clip}, yet these methods required partial knowledge of the generative processes during training. 
Additional works \citep{ojha2023towards, sha2023fake, cozzolino2023raising} leverage feature representations from large pre-trained models to enhance performance in data-limited settings. Still, some reliance on generated data or familiarity with generative techniques persists. 

\begin{figure*}[ht]
    \centering\includegraphics[width=0.8\textwidth]{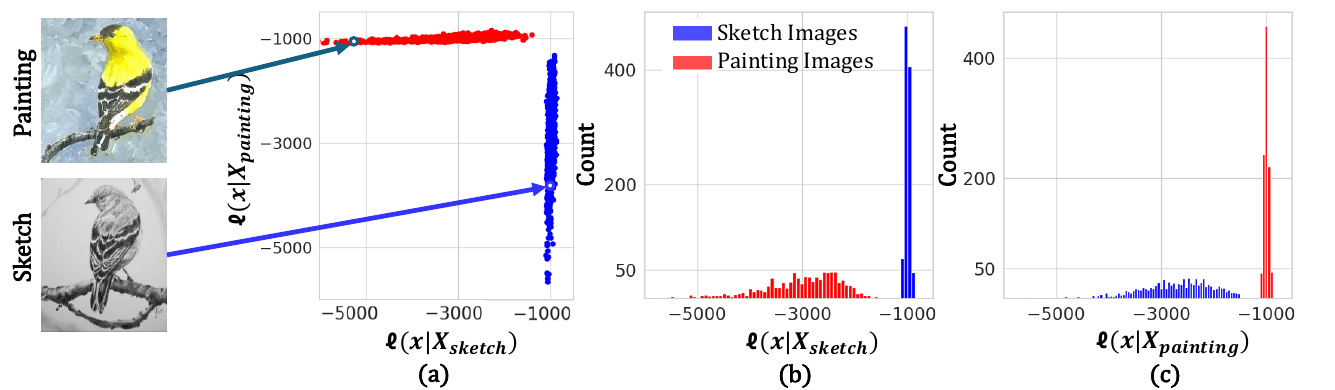}
    \caption{\textbf{Conditional likelihood across domains.} Despite semantic similarity, ``sketch'' and ``painting'' images from ImageNet-R yield separable distributions under our log-likelihood approximation. (a) 2D scatter plot of likelihoods conditioned on each domain. (b, c) Histograms show that each domain assigns higher likelihoods to its own images, with lower and more variable scores for the other.}

    \label{fig: conditional likelihood example}
\end{figure*}

Zero-shot detection strategies aim to identify generated images without prior exposure to any particular generative models or synthetic datasets. 
Existing zero-shot detection methods typically analyze an image's robustness to a specific transformation (\eg denoising, compression), and compare the original and transformed images to assess authenticity. AEROBLADE \cite{ricker2024aeroblade} employs a pre-trained auto-encoder, using its reconstruction error as a detection criterion, while RIGID \cite{he2024rigid} evaluates the similarity between an image and its noise-perturbed counterpart. Other approaches explore alternative transformations: ZED \cite{cozzolino2024zero} estimates pixel-wise probability distributions via lossless encoding; Manifold-Bias \cite{brokman2025manifold} uses differential measures in the diffusion model's probability space, this approach also generalizes to the memorization task \cite{brokman2025identifying, brokman2025tracking}.

While effective in a general setting, these approaches implicitly rely on approximate probability functions that can be unstable and sensitive to variations in generative models. Moreover, as shown in our experiments, their performance degrades significantly in domain-specific scenarios, highlighting a key shortcoming in generalization across real-world image distributions. In contrast, our method explicitly models the distribution of real images and offers inherent flexibility to adapt across domains, resulting in a more robust and versatile zero-shot detection solution.


\textbf{Likelihood approximation for images.}  
Estimating the likelihood of images is a fundamental problem with wide-ranging applications in computer vision \citep{yu2018generative, tian2020deep, goyal2020image, elharrouss2020image, li2022srdiff}. The likelihood of an image refers to the probability density function that describes how probable a given image is under a learned or assumed distribution of natural images. The early methods relied on assumptions about the smoothness of natural images \citep{geman1984stochastic} and the statistical distribution of image patches \citep{zoran2011learning}. The advent of generative models, including GANs \citep{goodfellow2020generative} and Diffusion Models \citep{song2020denoising}, has significantly improved image synthesis by learning implicit representations of the likelihood. However, these models do not provide direct access to the likelihood function itself. Recently, \cite{betser2025whitened} proposed leveraging CLIP \cite{radford2021learning} to obtain a likelihood approximation through the whitening transform. Following this, we introduce a conditional likelihood approximation and apply it to the task of generated image detection.

\section{Conditional Likelihood Approximation}

We follow \cite{betser2025whitened}, which introduced the use of whitened CLIP (W-CLIP) for likelihood estimation. Here, we extend this general likelihood estimation framework to a conditional likelihood estimation.

\subsection{Global CLIP likelihood (recap)}

\textbf{Notations:}
Let $\mathcal{X}=\{x_1,\dots,x_N\}$ be $N$ vectors in $\mathbb{R}^{d}$, and let $\mathbf{X}\in\mathbb{R}^{d\times N}$ denote their column-stacked matrix, with mean $\mu = \frac{1}{N} \sum_{i=1}^{N} x_i$. The centered vectors are $\hat{x}_i = x_i - \mu$, and let $\mathbf{\hat{X}}\in\mathbb{R}^{d\times N}$ be the column-stacked matrix of centered vectors. The empirical covariance matrix is $\mathbf{\Sigma} = \frac{1}{N} \mathbf{\hat{X}} \mathbf{\hat{X}}^\top$.

\textbf{Whitening Transform:}  
We compute a whitening matrix \( W \in \mathbb{R}^{d \times d} \) satisfying \( W^\top W = \mathbf{\Sigma}^{-1} \). Using PCA, we diagonalize \( \mathbf{\Sigma} \) as \( \mathbf{\Sigma} = V \Lambda V^\top \), where \( \Lambda \) contains eigenvalues and \( V \) contains eigenvectors. The whitening matrix is \( W = \Lambda^{-\frac{1}{2}} V^\top \). A vector \( x \) is whitened via \( y = W \hat{x} \), and the whitened matrix is \( \mathbf{Y} = W \mathbf{\hat{X}} \). The whitened vectors are isotropic with zero mean and identity covariance.

\textbf{Likelihood Approximation:}  
Prior work suggests CLIP’s image and text latent spaces are approximately disjoint \cite{liang2022mind, levi2024double}, allowing each modality to be modeled independently. When whitening statistics are estimated on a broad real-image corpus (e.g., MS-COCO val), the whitened CLIP embeddings have identity covariance by construction; empirical tests on MS-COCO (Anderson–Darling \cite{anderson1954test}, D’Agostino–Pearson \cite{d1973tests}) further indicate that the whitened coordinates are approximately i.i.d.\ standard normal. The log-likelihood of a whitened vector \( y \) is:
\begin{equation}
\ell(y) = -\frac{1}{2} \left( d \log (2 \pi) + |y|^2 \right),
\label{eq:isotropic_log_likelihood}
\end{equation}
where \( |y|^2 = y^\top y \). Given an embedding \( x \), the whitened vector \( y = W \hat{x} \) provides a direct likelihood approximation.

\begin{figure*}
    \centering
    \includegraphics[width=0.85\linewidth]{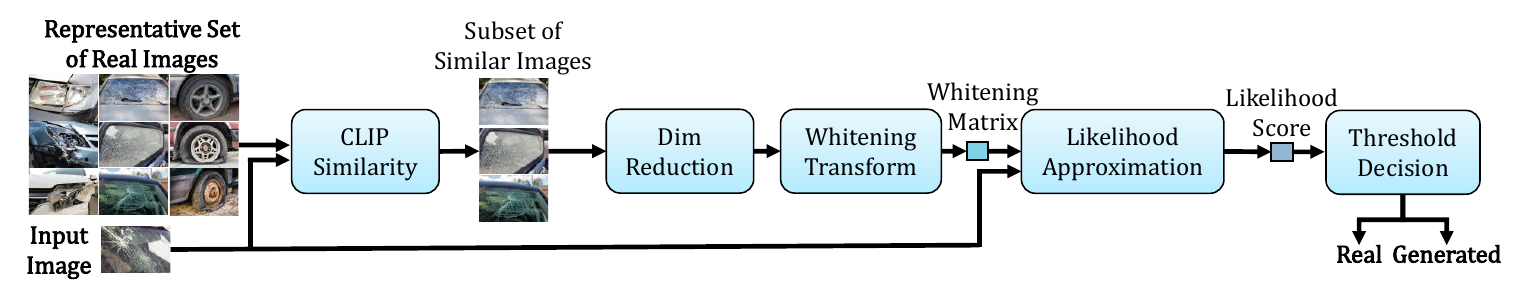}
    \caption{\textbf{Method flowchart.} 
    The input image is embedded using CLIP, followed by a selection of a subset $k$ of similar image embeddings from the representative set ($X$). A whitening matrix ($W$), with reduced dimensions ($m$), is computed using this subset ($X_k$). The whitening matrix is then applied to calculate a likelihood score (\cref{eq:conditioned_log_likelihood}). The final decision is determined based on an empirical threshold.}

    \label{fig: method flow}
\end{figure*}

\textbf{Global likelihood gap.}
Contrastive training pulls semantically similar images together and pushes dissimilar ones apart along \emph{directions} in the embedding space \cite{saunshi2019theoretical}. The score in \cref{eq:isotropic_log_likelihood}, however, depends only on the squared norm $\|W\hat{x}\|^2$, discarding directional information. As a result, distinct domains that occupy different subspaces but lie at comparable radii from the origin can appear similarly likely under a \emph{global} surrogate. This motivates conditioning the covariance on domain evidence to assess domain-relevant directional structure.

\subsection{Conditional likelihood}
\label{sec: conditional likelihood}

Building on the global likelihood model, we formulate a \emph{family of domain-conditional} likelihoods by estimating whitening on domain-specific data and scoring within the domain’s dominant subspace. We also verify approximate normality on small ($\sim$1K) domain sets, empirically justifying the conditional likelihood estimation.

\textbf{Theory.}
Let the global likelihood follow \cref{eq:isotropic_log_likelihood}, and let \(y(j)\) denote the $j$-th coordinate of a whitened vector \(y\in\mathbb{R}^d\). Conditioning on a linear constraint (e.g., \(y(j)=c\)) reduces the effective dimension by one and yields a standard normal on a \(d-1\) dimensional subspace; more generally, conditioning on $l$ independent constraints restricts the support to a $(d-l)$-dimensional subspace with standard normal coordinates on that subspace. In practice, domain evidence induces \emph{low variance} along some directions; estimating the covariance on domain-specific data and restricting to the top-$m$ eigenvectors (with $m\le d$) recovers an isotropic model on the domain’s dominant subspace.


\textbf{Domain taxonomy.} Each image can belong to multiple, possibly overlapping \textit{domains}, i.e., sets of images sharing characteristics (e.g., an image of a yellow bus in a city belongs to real-world, transportation, color, and urban domains).


\textbf{Model.}
Given a domain $\mathcal{D}$ occupies an (unknown) subspace of $m=d-l$ dominant dimensions, with the remaining $l$ directions exhibiting low variance. Given domain evidence $X=\{x_i\}\subset\mathcal{D}$, we compute the conditional whitening matrix $W(X,m)\in\mathbb{R}^{m\times d}$ using the top-$m$ eigenvectors. The conditional log-likelihood is then:

\begin{equation}
\ell(x \mid X,m)
\;=\;
-\tfrac{1}{2}\!\left(m\log(2\pi)\;+\;\big\|\,W(X,m)\,\hat{x}\,\big\|^2\right).
\label{eq:conditioned_log_likelihood}
\end{equation}



\textbf{Empirical validation.}
A necessary precondition for conditional likelihood is that, after whitening on domain-specific evidence, coordinates are approximately Gaussian on the domain’s dominant subspace. We validate this on small ($\sim$1K) ImageNet-R \cite{hendrycks2021many} domains (\emph{sketch}, \emph{painting}): normality tests on whitened coordinates pass clearly (Supp.~Sec.~B), and cross-domain (e.g., $W_{\text{sketch}}$ applied to \emph{painting} and vice versa) preserves acceptable Gaussianity while inducing systematic likelihood shifts.
As shown in \cref{fig: conditional likelihood example}, the 2D scatter and 1D histograms exhibit clear cross-domain separation, demonstrating robust distributional distinctions even between semantically similar images. As shown in Supp.~Sec.~B, Fig.~3b, retaining all $d$ dimensions yields substantial inter-coordinate correlations, violating the isotropic/i.i.d.\ assumption underlying \cref{eq:conditioned_log_likelihood}. Consequently, restricting the whitened space to a lower-dimensional subspace (top-$m$ components) is essential.

\noindent\textbf{Conditional vs.\ global.}
Whereas W-CLIP estimates a single \emph{global} surrogate on a broad real-image corpus, our formulation provides a \emph{family of domain-conditional} likelihoods by using data limited to a specific domain. This shift is conceptually distinct and empirically necessary to avoid global-surrogate failure modes, and it directly enables our zero-shot detector (Sec.~4). Statistical checks (Supp.~Sec~B) and a side-by-side comparison of global vs.\ conditional models and performance are provided in Supp.~Sec~J.

\begin{table*}[htbp]
\centering
\rowcolors{1}{white}{gray!20}
\resizebox{\textwidth}{!}{%
\begin{tabular}{|l|cccc|cccc|cccc|cccc|cccc|}
\toprule
\multirow{2}{*}{\textbf{\shortstack{Generative Model}}} & \multicolumn{4}{c|}{\textbf{AEROBLADE} \cite{ricker2024aeroblade}} & \multicolumn{4}{c|}{\textbf{RIGID} \cite{he2024rigid}} & \multicolumn{4}{c|}{\textbf{ZED} \cite{cozzolino2024zero}} & \multicolumn{4}{c|}{\textbf{Manifold Bias} \cite{brokman2025manifold}} & \multicolumn{4}{c|}{\textbf{CLIDE (ours)}} \\
\cmidrule(lr){2-5} \cmidrule(lr){6-9} \cmidrule(lr){10-13} \cmidrule(lr){14-17} \cmidrule(lr){18-21}  
& AUC $\uparrow$ & AP $\uparrow$ & F1 $\uparrow$ & Acc $\uparrow$  
& AUC $\uparrow$ & AP $\uparrow$ & F1 $\uparrow$ & Acc $\uparrow$  
& AUC $\uparrow$ & AP $\uparrow$ & F1 $\uparrow$ & Acc $\uparrow$  
& AUC $\uparrow$ & AP $\uparrow$ & F1 $\uparrow$ & Acc $\uparrow$  
& AUC $\uparrow$ & AP $\uparrow$ & F1 $\uparrow$ & Acc $\uparrow$  \\
\midrule
ProGan \cite{karras2017progressive} & 0.54 & 0.49 & 0.65 & 0.54 & 0.45 & 0.46 & 0.16 & 0.48 & 0.74 & 0.75 & 0.7 & 0.64 & \textbf{0.96} & \textbf{0.97} & \textbf{0.88} & \textbf{0.88} & 0.95 &	\textbf{0.96} &	0.87 &	\textbf{0.88} \\
StyleGan \cite{karras2019style} & 0.58 & 0.59 & 0.63 & 0.51 & 0.71 & 0.72 & 0.56 & 0.65 & 0.75 & \textbf{0.77} & 0.71 & 0.65 & 0.68 & 0.74 & 0.61 & 0.68 & \textbf{0.76} &	0.71 &	\textbf{0.75} &	\textbf{0.73} \\
StyleGan2 \cite{karras2020analyzing} & 0.74 & 0.76 & 0.68 & 0.61 & 0.52 & 0.54 & 0.31 & 0.53 & 0.75 & 0.8 & 0.68 & 0.61 & 0.62 & 0.66 & 0.48 & 0.6 & \textbf{0.8} & \textbf{0.82} & \textbf{0.73} & \textbf{0.68} \\
BigGAN \cite{brock2018large} & 0.65 & 0.61 & 0.68 & 0.6 & 0.62 & 0.63 & 0.42 & 0.57 & 0.45 & 0.47 & 0.63 & 0.5 & 0.9 & 0.9 & 0.8 & 0.81 & \textbf{0.96} & \textbf{0.96} & \textbf{0.87} & \textbf{0.88} \\
GauGan \cite{park2019semantic} & 0.7 & 0.71 & 0.68 & 0.6 & 0.32 & 0.39 & 0.1 & 0.45 & 0.33 & 0.41 & 0.6 & 0.44 & \textbf{0.98} & \textbf{0.98} & \textbf{0.91} & \textbf{0.9} & 0.83 & 0.85 & 0.76 & 0.73 \\
CycleGan \cite{zhu2017unpaired} & 0.76 & 0.73 & 0.73 & 0.69 & 0.34 & 0.39 & 0.05 & 0.43 & 0.22 & 0.38 & 0.6 & 0.43 & 0.97 & 0.97 & 0.89 & 0.9 & \textbf{0.98} & \textbf{0.98} & \textbf{0.9} & \textbf{0.91} \\
CRN \cite{chen2017photographic} & 0.34 & 0.51 & 0.48 & 0.59 & 0.25 & 0.36 & 0.01 & 0.42 & \textbf{0.99} & \textbf{0.99} & \textbf{0.92} & \textbf{0.92} & 0.93 & 0.9 & 0.88 & 0.88 & 0.98 & \textbf{0.99} & 0.91 & \textbf{0.92} \\
SD V1.4 \cite{rombach2022high} & 0.47 & 0.47 & 0.63 & 0.51 & 0.79 & 0.77 & 0.63 & 0.69 & 0.62 & 0.65 & 0.65 & 0.54 & 0.71 & 0.62 & 0.4 & 0.57 & \textbf{0.9} & \textbf{0.91} & \textbf{0.82} & \textbf{0.82} \\
SD V1.5 \cite{rombach2022high} & 0.49 & 0.49 & 0.64 & 0.52 & 0.77 & 0.76 & 0.63 & 0.68 & 0.61 & 0.63 & 0.65 & 0.54 & 0.72 & 0.64 & 0.44 & 0.58 & \textbf{0.9} & \textbf{0.91} & \textbf{0.83} & \textbf{0.82} \\
Guided DM \cite{dhariwal2021diffusion} & 0.55 & 0.49 & 0.66 & 0.57 & 0.51 & 0.51 & 0.25 & 0.5  & 0.58 & 0.55 & 0.68 & 0.59 & 0.8 & 0.8 & 0.66 & 0.7 & \textbf{0.87} & \textbf{0.87} & \textbf{0.79} & \textbf{0.78}\\
LDM 100 \cite{rombach2022high} & 0.48 & 0.48 & 0.62 & 0.48 & 0.51 & 0.51 & 0.25 & 0.5 & 0.52 & 0.56 & 0.63 & 0.49 & 0.84 & 0.88 & 0.78 & 0.79 & \textbf{0.92} & \textbf{0.93} & \textbf{0.84} & \textbf{0.85} \\
LDM 200 \cite{rombach2022high} & 0.47 & 0.48 & 0.62 & 0.48 & 0.51 & 0.51 & 0.25 & 0.5 & 0.53 & 0.56 & 0.63 & 0.5 & 0.85 & 0.88 & 0.79 & 0.8 & \textbf{0.92} & \textbf{0.93} & \textbf{0.85} & \textbf{0.85} \\
Glide 50 27 \cite{nichol2021glide} & 0.06 & 0.32 & 0.59 & 0.42 & 0.51 & 0.51 & 0.25 & 0.5 &\textbf{0.98} & \textbf{0.98} & \textbf{0.91} & \textbf{0.92} & 0.93 & 0.94 & 0.86 & 0.86 & 0.91 & 0.92 & 0.84 & 0.84 \\
Glide 100 27 \cite{nichol2021glide} &  0.08 & 0.32 & 0.6 & 0.43 & 0.51 & 0.51 & 0.25 & 0.5 & \textbf{0.98} & \textbf{0.98} & \textbf{0.91} & \textbf{0.92} & 0.92 & 0.93 & 0.84 & 0.84 & 0.91 & 0.92 & 0.83 & 0.83\\
Glide 100 10 \cite{nichol2021glide} & 0.04 & 0.31 & 0.59 & 0.42 & 0.51 & 0.51 & 0.25 & 0.5 & \textbf{0.98} & \textbf{0.98} & \textbf{0.92} & \textbf{0.92} & 0.93 & 0.93 & 0.85 & 0.85 & 0.91 & 0.92 & 0.83 & 0.83 \\
ADM \cite{dhariwal2021diffusion} & 0.45 & 0.43 & 0.65 & 0.56 & 0.56 & 0.57 & 0.39 & 0.56 & 0.47 & 0.46 & 0.65 & 0.55 & 0.62 & 0.57 & 0.32 & 0.53 & \textbf{0.89} & \textbf{0.88} & \textbf{0.84} & \textbf{0.84} \\
DALL·E 3 \cite{betker2023improving} & 0.35 & 0.41 & 0.61 & 0.45 & 0.51 & 0.51 & 0.25 & 0.5 & 0.45 & 0.5 & 0.62 & 0.47 & 0.82 & 0.84 & 0.71 & 0.74 & \textbf{0.96} & \textbf{0.96} & \textbf{0.88} & \textbf{0.89} \\
Midjourney \cite{midjourney} & 0.29 & 0.37 & 0.6 & 0.43 & 0.65 & 0.66 & 0.48 & 0.6 & \textbf{0.92} & \textbf{0.93} & 0.84 & 0.84 & 0.56 & 0.53 & 0.27 & 0.51 & 0.9 & 0.84 & \textbf{0.85} & \textbf{0.85} \\
VDQM \cite{gu2022vector} & 0.51 & 0.47 & 0.66 & 0.57 & 0.53 & 0.54 & 0.32 & 0.53 & 0.46 & 0.46 & 0.63 & 0.49 & 0.83 & 0.83 & 0.7 & 0.73 & \textbf{0.92} & \textbf{0.93} & \textbf{0.84} & \textbf{0.84} \\
Wukong \cite{mindspore2024wukong} & 0.48 & 0.47 & 0.62 & 0.49 & 0.63 & 0.6 & 0.37 & 0.55 & 0.39 & 0.45 & 0.6 & 0.44 & 0.73 & 0.69 & 0.5 & 0.62 & \textbf{0.95} & \textbf{0.96} & \textbf{0.88} & \textbf{0.88} \\
\midrule
\rowcolor{gray!40}
\textbf{All} & 0.52 & 0.48 & 0.64 & 0.53 & 0.51 & 0.53 & 0.28 & 0.52 & 0.69 & 0.66 & 0.69 & 0.62 & 0.85 & 0.88 & 0.76 & 0.78 & \textbf{0.91} & \textbf{0.91} & \textbf{0.84} & \textbf{0.84} \\
\bottomrule
\end{tabular}%
}
\caption{\textbf{Evaluation of detection on general images.} Our method achieves the highest overall performance across all models combined and outperforms competing methods on most generative models. The best result for each metric in each row is highlighted in bold.}
\label{tab:general images}
\end{table*}


\section{Method: CLIDE}


We introduce \textbf{CLIDE} (\textbf{C}onditional \textbf{L}kelihood generated \textbf{I}mage \textbf{DE}tector), a zero-shot generated image detection method based on the proposed conditional likelihood approximation, conditioning on \textit{real images} to distinguish generated content. An overview is given in \cref{fig: method flow} and in the high-level \cref{alg: method}. A detailed algorithm is in Supp. Sec. A. In the following, we outline the key components of our approach.

\begin{algorithm}[ht]
\caption{CLIDE - high level}
\begin{algorithmic}[1]
\Inputs
$I$ (input image), $X$ (reference set), $k$ (neighbors), $m$ (components), $th$ (threshold)
\Outputs
Likelihood score $\ell(I \mid X, k, m)$, label $C$


\State \textbf{Step 1:} Embed the input image to obtain $x$, and compute cosine similarities with the embeddings in $X^{d \times N}$.
\State \textbf{Step 2:} Select the top-$k$ most similar embeddings - $X_k$.
\State \textbf{Step 3:} Compute mean vector $\mu(X_k)$ and whitening matrix $W(X_k, m)$ using top-$m$ eigenvectors.
\State \textbf{Step 4:} Compute likelihood $\ell(I \mid X, k, m)$ and compare to threshold $th$ to assign label $C$.
\end{algorithmic}
\label{alg: method}
\end{algorithm}

\noindent \textbf{Motivation.}  
CLIP embeddings encode semantic information \cite{radford2021learning}, which can bias likelihood estimates. For example, zebras tend to have higher likelihoods than plates of food, making a generated zebra potentially more likely than a real plate (Supp. Sec. C, Fig. 4). CLIP also captures domain-specific traits, as shown in \cref{fig: conditional likelihood example}, where conditioning shifts the distribution of similar images (\eg birds on branches) across domains. To improve detection, we aim to minimize semantic bias while enhancing domain-specific separation.





\noindent \textbf{Whitening Data Selection.}  
Distinguishing real from high-quality generated images is subtle. We address this with adaptive whitening. Given a set \( X \) of real images, we select the \( k \) most similar images to the input based on cosine similarity of CLIP embeddings. The whitening matrix is computed using this subset, \( W(X_k, m) \), enabling the conditioned likelihood to adapt to the local semantic subspace.

\noindent \textbf{Dimensionality Reduction.} 
Conditioned likelihood operates in a subspace of the embedding space (\cref{sec: conditional likelihood}), defined by the selected samples for whitening. Additionally, when using a small number of samples for whitening, some eigenvalues approach zero, causing instability. Dimensionality reduction mitigates these issues and  enhances both performance and efficiency (Supp. Sect. B, E, Figs. 3, 6).

\noindent \textbf{Domain-adaptation.}  Our method can target a specific domain and distinguish real from generated content within it. As in general domain adaptation methods, a small set of images from the target domain is needed to support detection. This set serves as the representative set \( X \) for conditional likelihood estimation. In \cref{sec: ablation}, we examine the method’s robustness to different choices of \( X \), showing many real-world datasets yield similar results in the general case. For domain specific detection, target samples must be included in \( X \), but do not need to dominate it; mixed domain sets still support high detection performance across domains.

By integrating these steps into the computation of the likelihood score, we can directly utilize it as a differentiation criterion between real and generated images. Our method has unique characteristics distinguishing it compared to existing zero-shot detection methods:
\begin{itemize}
    \item \textbf{Real image likelihood.} First method utilizing a likelihood estimation for the generated image detection task.
    \item \textbf{Domain adaptability.} Using sample images for the conditional likelihood enables a domain-adaptive approach.
    \item \textbf{Explicit probabilistic computation.} Other methods implicitly try to find a probabilistic differentiation between real and generated images. Our method explicitly computes the probabilities, yielding a stable distinction.

\end{itemize}

\begin{table*}[htbp]
\centering
\rowcolors{1}{white}{gray!20}
\resizebox{\textwidth}{!}{%
\begin{tabular}{|l|cccc|cccc|cccc|cccc|cccc|}
\toprule
\multirow{2}{*}{\textbf{\shortstack{Generative Model}}} & \multicolumn{4}{c|}{\textbf{AEROBLADE} \cite{ricker2024aeroblade}} & \multicolumn{4}{c|}{\textbf{RIGID} \cite{he2024rigid}} & \multicolumn{4}{c|}{\textbf{ZED} \cite{cozzolino2024zero}} & \multicolumn{4}{c|}{\textbf{Manifold Bias} \cite{brokman2025manifold}} & \multicolumn{4}{c|}{\textbf{CLIDE (ours)}} \\
\cmidrule(lr){2-5} \cmidrule(lr){6-9} \cmidrule(lr){10-13} \cmidrule(lr){14-17} \cmidrule(lr){18-21}  
& AUC $\uparrow$ & AP $\uparrow$ & F1 $\uparrow$ & Acc $\uparrow$  
& AUC $\uparrow$ & AP $\uparrow$ & F1 $\uparrow$ & Acc $\uparrow$  
& AUC $\uparrow$ & AP $\uparrow$ & F1 $\uparrow$ & Acc $\uparrow$  
& AUC $\uparrow$ & AP $\uparrow$ & F1 $\uparrow$ & Acc $\uparrow$  
& AUC $\uparrow$ & AP $\uparrow$ & F1 $\uparrow$ & Acc $\uparrow$  \\
\midrule
\multicolumn{21}{|c|}{\textbf{Artistic Image Domain}} \\
\midrule
StyleGan3 \cite{karras2021alias} & 0.81 & 0.84 & 0.72 & 0.69 & 0.32 & 0.39 & 0.06 & 0.46 & 0.38 & 0.41 & 0.63 & 0.5 & 0.96 & 0.96 & 0.9 & 0.9 & \textbf{0.98} & \textbf{0.99} & \textbf{0.93} &  \textbf{0.93} \\
SD2.1 \cite{rombach2022high} & 0.57 & 0.6 & 0.61 & 0.48 & 0.77 & 0.75 & 0.52 & 0.64 & 0.17 & 0.34 & 0.61 & 0.44 & 0.91 & 0.9 & 0.83 & 0.84 & \textbf{0.98} & \textbf{0.98} & \textbf{0.92} & \textbf{0.93} \\
SDXL \cite{podell2023sdxl} & 0.62 & 0.69 & 0.6 & 0.46 & 0.64 & 0.6 & 0.3 & 0.54 & 0.33 & 0.43 & 0.61 & 0.44 & 0.83 & 0.81 & 0.71 & 0.75 & \textbf{0.97} & \textbf{0.98} & \textbf{0.91} & \textbf{0.91} \\
AnimagineXL \cite{animagine} & 0.66 & 0.75 & 0.6 & 0.47 & 0.39 & 0.42 & 0.08 & 0.46 & 0.45 & 0.59 & 0.61 & 0.44 & 0.28 & 0.37 & 0.02 & 0.44 & \textbf{0.76} & \textbf{0.83} & \textbf{0.66} & \textbf{0.55} \\
\midrule
\rowcolor{gray!40}
All & 0.65 & 0.7 & 0.62 & 0.52 & 0.53 & 0.55 & 0.26 & 0.53 & 0.33 & 0.41 & 0.61 & 0.45 & 0.74 & 0.8 & 0.69 & 0.73 & \textbf{0.92} & \textbf{0.92} & \textbf{0.83} & \textbf{0.82} \\

\midrule
\multicolumn{21}{|c|}{\textbf{Damaged cars Image Domain}}
\\
\midrule
Kandinsky2.1 \cite{razzhigaev2023kandinsky} & 0.29 & 0.38 & 0.6& 0.43 & 0.3 & 0.38 & 0.06 & 0.44 & 0.68 & 0.66 & 0.68 & 0.61 & 0.6& 0.64 & 0.48 & 0.61 & \textbf{0.97} & \textbf{0.98} & \textbf{0.93} &  \textbf{0.93} \\
SDXL \cite{podell2023sdxl} & 0.34 & 0.4& 0.61 & 0.45 & 0.3& 0.38 & 0.06 & 0.44 & 0.83 & 0.85 & 0.76 & 0.73 & 0.32 & 0.42 & 0.15 & 0.49 & \textbf{0.97} & \textbf{0.97} & \textbf{0.91} & \textbf{0.91} \\
DALL·E 3 \cite{betker2023improving}& 0.39 & 0.45 & 0.6 & 0.44 & 0.4 & 0.42 & 0.11 & 0.42 & 0.15 & 0.33 & 0.59 & 0.42 & 0.14 & 0.33 & 0.03 & 0.44 & \textbf{0.99} & \textbf{0.99} & \textbf{0.93} & \textbf{0.93} \\
Midjourney \cite{midjourney} & 0.46 & 0.47 & 0.62 & 0.47 & 0.51 & 0.5 & 0.23 & 0.5 & 0.78 & 0.75 & 0.74 & 0.71 & 0.4 & 0.44 & 0.13 & 0.48 & \textbf{0.93} & \textbf{0.94} & \textbf{0.87} & \textbf{0.86} \\
UnCLIP \cite{ramesh2022hierarchical} & 0.22 & 0.35 & 0.6 & 0.43 & 0.69 & 0.67 & 0.46 & 0.6 & 0.31 & 0.38 & 0.61 & 0.46 & 0.69 & 0.68 & 0.48 & 0.61 & \textbf{0.8} & \textbf{0.8} & \textbf{0.72} & \textbf{0.67} \\
\midrule
\rowcolor{gray!40}
All & 0.33 & 0.4 & 0.61 & 0.44 & 0.45 & 0.48 & 0.21 & 0.49 & 0.61 & 0.54 & 0.68 & 0.61 & 0.47 & 0.53 & 0.3 & 0.53 & \textbf{0.92} & \textbf{0.92} & \textbf{0.85} & \textbf{0.84} \\

\midrule
\rowcolor{gray!20}
\multicolumn{21}{|c|}{\textbf{Invoice Image Domain}}
\\
\midrule
\rowcolor{gray!40}
GPT-Image-1 \cite{openai2025gptimage1} & 0.53 & 0.62 & 0.56 & 0.47 & 0.57 & 0.67 & 0.12 & 0.53 & 0.6 & 0.56 & 0.65 & 0.54 & 0.61 & 0.58 & 0.55 & 0.62 & \textbf{0.91} & \textbf{0.92} & \textbf{0.82} &  \textbf{0.82} \\
\bottomrule
\end{tabular}%
}
\caption{\textbf{Detection performance on domain-specific images.} Our method consistently maintains high performance, while others show significant drops. It outperforms all baselines across all three domains - both per model and when tested on all models combined - using a single model for the invoice domain due to the lack of other high-quality sources. Best results per row are in bold.}

\label{tab:domain_images}
\end{table*}

\section{Evaluation}
\begin{figure}[ht]
    \centering\includegraphics[width=0.45\textwidth]{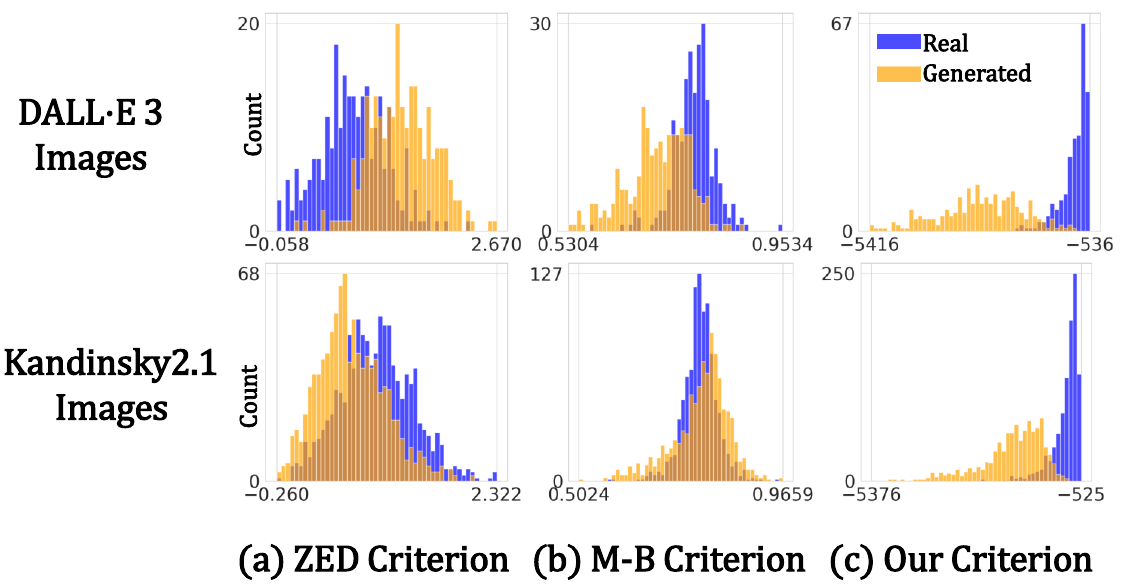}
    \caption{\textbf{Inconsistent separation direction - ``flipped classification''.} ZED (a) and Manifold Bias (b) show inconsistent behavior across generative models, with real and generated scores flipping. Our method (c) yields consistent results, assigning higher likelihoods to real images. Example from the damaged cars domain.}
 
    \label{fig: opp hists}
\end{figure}

We evaluate in two setups. First, a \textit{general} setting using a diverse image benchmark to assess detection ability and compare against leading zero-shot methods. Second, we introduce a \textit{domain-specific detection} task that targets three concrete domains-artistic images, damaged cars, and invoice documents. Both real and generated content come from a narrow distribution. This setting is underexplored yet practically critical (e.g., fraud detection). Across these domains, \emph{all} existing zero-shot detectors degrade sharply (often with AUC $<\!0.6$ and flipped decisions; see Tabs.~\ref{tab:domain_images} and Fig.~\ref{fig: opp hists}), whereas our conditional likelihood maintains high accuracy (AUC $\gtrsim 0.9$). We additionally provide the first synthetic benchmarks for damaged cars and invoices. This dual evaluation highlights both standard zero-shot performance and, crucially, CLIDE’s unique strength in domain-adaptive detection.

\textbf{Implementation Details.}  
We set \( k = 500 \) and \( m = 400 \) (real samples for whitening and dimensions) across all domains (see \cref{sec: ablation} for details). We use the CLIP ViT-L/14 model \cite{radford2021learning}. Our full code is provided in the Supplementary Material. For AEROBLADE and Manifold Bias, we use official implementations. As no official code exists for RIGID and ZED, we self-implement them. For RIGID, we use the DINO model \cite{caron2021emerging}; for ZED, we apply the lossless compression method SReC \cite{cao2020lossless}. ZED proposes four criteria; we report the best-performing one in the general image domain (\( D_0 \)). Results for the other criteria are provided in Supp. Sec. H, Tabs. 2 and 3.

\textbf{Metrics.}  
We evaluate each detection method using four metrics. To measure the separation between real and generated image distributions, we use two standard metrics: Area Under the Curve (AUC) and Average Precision (AP). Classification performance is further assessed using F1 score and accuracy. The binary classification threshold is set for all methods as \( th = \text{mean}(C_{th}) + \text{std}(C_{th}) \), where \( C_{th} \) are the criterion values of a set of real images.


\begin{figure*}[htbp]
    \centering\includegraphics[width=0.85\textwidth]{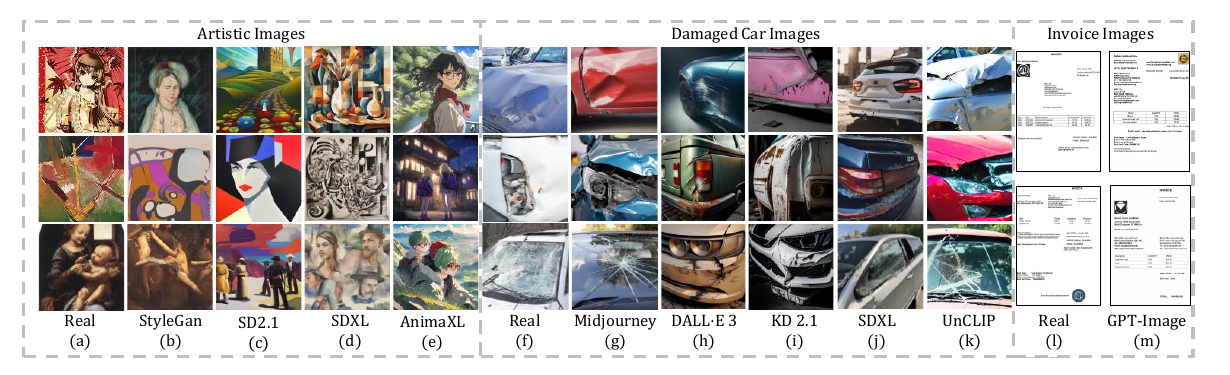}
    \caption{\textbf{Example images.} Examples of real images (a \cite{wikiart, danbooru2021}, f \cite{wang2023cardd}, m \cite{wiF0n2025invoiceXpert}) and images generated using different generative models from different domains. Artistic generated images (b-e) are taken from \cite{boychev2025imaginet} and all other generated images are generated by us.}
    \label{fig: domain example}
\end{figure*}

\subsection{General image detection}
\textbf{Dataset.} In the general image domain, we follow \cite{brokman2025manifold} and combine three large-scale datasets \citep{wang2020cnn, ojha2023towards, zhu2023genimage}, containing 100K generated images from GANs, diffusion models, and commercial tools. A full list of the 20 generative models is provided in \cref{tab:general images}. We use the real images from \citep{wang2020cnn, ojha2023towards}, originally sourced from LAION \cite{schuhmann2021laion}. As there are only 50K real images from LAION, we sample an additional 50K from the MS-COCO training set to form the 100K real subset used in the "all" row of \cref{tab:general images}, totaling 200K images for evaluation. Each generative model contributes at least 1K generated samples, which are evaluated against an equal number of randomly selected real images from LAION, consistently applied across all detection methods. Following \cite{betser2025whitened}, we use 5K MS-COCO validation images for threshold calibration for all methods, and as the representative set for our method. An ablation appears in \cref{sec: ablation}.

\textbf{Results.} The last row in \cref{tab:general images} shows that our method outperforms all other zero-shot detection approaches. 
Our method also achieves the best results per generative model in most cases. While ZED or Manifold Bias surpass our approach for certain models, our results remain competitive. 
Notably, our proposed method consistently yields high performance across all models, with AUC dropping below 0.8 only for StyleGAN (0.76). In contrast, all other detection methods exhibit AUC values below 0.6 for certain generative models. We recall that an AUC below 0.5 indicates a complete failure for binary classification, where
generated images are more likely to be classified as real ones. 
We refer to this phenomenon as ``flipped classification'' and illustrate it in \cref{fig: opp hists}. 
Since detection methods aim to maintain robustness across multiple models, this presents a significant limitation.

\subsection{Domain specific detection}
While our method performs well in general image detection, its key advantage is adaptability to new domains. To demonstrate this, we evaluate all methods across three specific domains. Existing approaches can only use domain-specific data for threshold calibration, whereas our method also leverages it for likelihood estimation. In each domain, we assume access to 1K real images (separate from the test set), used by all methods for calibration. Our method also uses them as the representative set.

\textbf{Artistic image domain.}
The generation of images in specific artistic styles is a well-studied area in image synthesis and style transfer \cite{gatys2016image, jing2019neural, wang2023stylediffusion,brokman2024montrage}. Given the diversity of artistic styles across different artists, this is a broad and varied domain of images. We use images from ImagiNet \cite{boychev2025imaginet}, a real and generated artistic image dataset. Synthetic images are generated using four different generative models (see \cref{tab:domain_images}). Examples of real and generated images are shown in \cref{fig: domain example}. Detection results for all methods are summarized in \cref{tab:domain_images}. In this domain, AnimaginXL \cite{animagine} presents the most challenging generative model, as it is specifically fine-tuned for generating images in a distinct artistic style (Anime). Our method outperforms all detection approaches across all generative models and in the combined evaluation (``all''). Furthermore, unlike competing methods, our performance remains consistent with the general image case (\cref{tab:general images}), whereas the strongest competitors in the general image domain (\cite{cozzolino2024zero}, \cite{brokman2025manifold}) exhibit a decline in performance.

\noindent \textbf{Damaged car image domain.}
Damaged car images form a niche but highly practical domain, relevant to insurance, rental, and second-hand vehicle markets \cite{wang2023cardd}. Prior work has focused on damage detection and classification \cite{patil2017deep, panboonyuen2023mars, hasan2024groundingcardd}, but progress has been hindered by the lack of public benchmarks, with early studies relying on private datasets. CarDD \cite{wang2023cardd} addressed this by releasing 4K real damaged car images; We use the training set (2,816 images) for evaluation and sample 1K calibration images from the validation and test splits. As no synthetic dataset exists for this domain, we generate the \emph{first publicly available set of synthetic damaged car images} (see \cref{fig: domain example}). The data is produced using two diffusion models, two commercial tools, and UnCLIP \cite{ramesh2022hierarchical}, which is conditioned on real images to enhance realism and detection difficulty. Additional dataset details and samples are in Supp. Sec. F.1; the full dataset will be released upon acceptance. Our method maintains high performance in this domain, comparable to results in general and artistic domains, while other methods degrade (see \cref{tab:domain_images}). A key issue is flipped classification (\cref{fig: opp hists}), where ZED and Manifold Bias often misclassify generated images as real. In contrast, our method remains stable, consistently achieving AUC scores above 0.5 across all models. While ZED performs well on some models, its overall zero-shot reliability is limited by this failure mode.

\begin{table}[ht]
\centering
\resizebox{\linewidth}{!}{%
\begin{tabular}{l|cccc}
\toprule
\textbf{Rep. Set Composition} & \textbf{LAION} & \textbf{MS-COCO} & \textbf{Flickr} & \textbf{Combined} \\
\midrule
MS-COCO & 0.92 & 0.92 & 0.89 & 0.91 \\
LAION & 0.9 & 0.86 & 0.85 & 0.87 \\
Flickr8k & 0.89 & 0.88 & 0.95 & 0.9 \\
ImageNet & 0.73 & 0.75 & 0.77 & 0.75 \\
MS-COCO + Flickr8k & 0.91 & 0.92 & 0.93 & 0.93 \\
LAION + ImageNet & 0.86 & 0.82 & 0.83 & 0.83 \\
All Sources Combined & 0.92 & 0.92 & 0.92 & 0.91 \\
\bottomrule
\end{tabular}%
}
\caption{\textbf{Real data sources.} 
AUC values. Different representative set sources and test set sources of real images.}
\label{tab:general_rep_set}
\end{table}

\noindent \textbf{Invoice image domain}
Invoice images form a critical domain, particularly relevant to financial institutions, e-commerce platforms, and document verification systems. While prior work has focused on information extraction and document classification \cite{denk2019bertgrid, xu2020layoutlm, appalaraju2021docformer}, little attention has been given to the detection of synthetically generated invoices. Existing datasets typically contain only real invoices, with no public benchmark for synthetic invoice detection. The generation of high-quality, text-rich invoice images was not feasible until the recent release of the GPT-Image-1 model \cite{openai2025gptimage1}, which enables the creation of visually and semantically realistic document images. For this study, we use real invoices from \cite{wiF0n2025invoiceXpert}, selecting 3K images for evaluation and another 1K for calibration. Synthetic images are generated using GPT-Image-1, each conditioned on a corresponding test image with slight random variations in elements such as price, product name, or date. We rely exclusively on GPT-Image-1, as it is currently the only model capable of producing high-quality document images for this task (see examples with other models in Supp. Sec. F.2, Fig. 14). Examples of real and generated invoices are shown in \cref{fig: domain example}, and additional details are provided in Supp. Sec. F.2. Our method demonstrates high detection performance in this domain, while other methods suffer significant performance drops (see \cref{tab:domain_images}). These results highlight the importance of domain adaptability and the limitations of existing zero-shot detection methods in complex domains.

\subsection{Robustness and efficiency analysis}
\label{sec: ablation}
\noindent \textbf{Representative set.}
We study the impact of the representative set's composition in both settings: general and domain-specific image detection.

\textit{General detection.} We evaluate four real-world image sources as representative sets: MS-COCO (5K images), LAION (10K, from \cite{wang2020cnn}), Flickr8k \cite{hodosh2013framing} (8K), ImageNet \cite{deng2009imagenet} (50K), and their combinations. Evaluation is performed on four fixed test sets from different sources, each containing 10K real and 10K generated images (except Flickr, with 3K each). As shown in \cref{tab:general_rep_set}, our method consistently achieves high AUC across all sources, except for ImageNet (0.73), likely due to its object-centric nature. 

\textit{Domain-specific detection.} We evaluate three domains (Cars, Art, Invoices), using 2K real and 2K generated images per domain, along with the general domain (10K each). We compare three representative set types: general-only, domain-only (1K images), and combined. As shown in \cref{tab:domain_rep_set}, general-only sets yield lower AUC scores on specific domains, while adding just 1K domain-specific samples raises AUC to 0.91–0.94. Combining general and domain-specific sources further improves robustness across domains. Notably, multi-domain representative sets achieve AUC $>0.9$ for all domains and maintain strong general-domain performance. These results demonstrate that our method is robust to representative set composition and adapts well to new domains with minimal supervision.


\begin{table}[ht]
\centering
\resizebox{\linewidth}{!}{%
\begin{tabular}{l|cccc}
\toprule
\textbf{Rep. Set Composition} & \textbf{General} & \textbf{Damaged Cars} & \textbf{Artistic} & \textbf{Invoice} \\
\midrule
General only & 0.92 & 0.56 & 0.77 & 0.34 \\
Cars Only & 0.53 & 0.93 & 0.57 & 0.8 \\
Art Only & 0.82 & 0.55 & 0.94 & 0.27 \\
Invoice Only & 0.55 & 0.69 & 0.6 & 0.91 \\
General + Cars & 0.9 & 0.93 & 0.62 & 0.8 \\
General + Invoice & 0.92 & 0.56 & 0.79 & 0.91 \\
Cars + Art + Invoice & 0.8 & 0.93 & 0.95 & 0.91 \\
All Domains Combined & 0.9 & 0.93 & 0.93 & 0.91 \\
\bottomrule
\end{tabular}%
}
\caption{\textbf{Representative set domain combinations.} AUC scores. Adding domain-specific samples improves in-domain performance; combined sets perform well across all domains.}
\label{tab:domain_rep_set}
\end{table}

\textbf{Parameter Ablation study.}
We conduct ablation studies to evaluate the impact of the parameters in our method. We analyze the effect of the number of whitening dimensions (\(m\)) and samples (\(k\)), finding optimal values around \(m = 300\)-\(400\) and \(k = m + 100\). We also compare global and per-image (local) whitening matrices, finding that local sampling yields slightly better performance, while the global setting offers improved efficiency with minimal loss. Overall, the method is robust to parameter selection, results are shown in Supp. Sec. E, Figs. 6–8.

\textbf{Robustness to image perturbations.}
We assess the robustness of our method under common image degradations, including JPEG compression, Gaussian blur, random resizing and cropping, noise injection, and color jittering. Perturbations are applied only to inference images, with the representative set kept unchanged, isolating robustness to input variation. Results (Supp. Sec. E.2, Figs. 9–11) show that our method maintains high accuracy (AUC \textgreater 0.8) across all perturbations. Only under extreme distortions - such as noise with std = 0.5 or JPEG quality = 1\% - does performance drop below 0.8 AUC.

\textbf{Efficiency Analysis.}  
Our method operates efficiently, achieving inference speeds comparable to ZED (0.25 seconds per image), while other methods, particularly AEROBLADE, exhibit significantly higher latency (\textgreater 4 seconds per image; see Supp. Sec. G, Tab. 4). In terms of memory consumption, our approach requires less RAM and GPU memory than most alternatives. A detailed analysis is provided in Supp. Sec. G. Overall, our method is lightweight and practical, making it well-suited for zero-shot detection.

\section{Conclusion}

We introduced a novel conditional likelihood approach for zero-shot detection of generated images, leveraging CLIP embeddings and a whitening-based likelihood approximation conditioned on real images. Unlike existing methods, our approach naturally adapts to domain-specific scenarios, where others often underperform. Through extensive experiments on a large-scale dataset of over 200K real and generated images, we showed that our method outperforms prior zero-shot techniques in the general detection task. We further validated its adaptability on three challenging domains: artistic images, damaged car images and invoice images.

To support domain-specific evaluation, we presented two new synthetic datasets: over 6K damaged car images generated using five different models, and 3K invoice images generated using a single high-quality model. Our method is also efficient and lightweight, achieving fast inference and low memory usage. It is robust to input perturbations, parameter choices, and representative set variations. These findings highlight the promise of likelihood-based zero-shot detection and suggest that conditional likelihood estimation may serve as a foundation for broader tasks in computer vision such as out-of-distribution detection, and domain-adaptive representation learning.

\subsection*{Acknowledgements}
We would like to acknowledge  
support by the Israel Science Foundation (Grant 1472/23) and by the Ministry of Innovation, Science and Technology (Grants No. 5074/22, 8801/25). We thank Shani Weiss for her assistance with generating the synthetic datasets.

{
    \small
    \bibliographystyle{ieeenat_fullname}
    \bibliography{main}
}

\end{document}

%% file: preamble.tex
\usepackage{algorithm}
\usepackage{algorithmicx}
\usepackage{algpseudocode}
\usepackage{amsmath}
\usepackage{multirow}
\usepackage{booktabs}   
\usepackage[table]{xcolor}
\usepackage{adjustbox}  
\usepackage{graphicx}   
